\documentclass[a4paper,10pt]{article}

\usepackage[utf8]{inputenc}
\usepackage{authblk}

\usepackage{graphicx}
\usepackage{url}
\usepackage{gensymb}

\title{Extending Continuum Models for Atom Probe Simulation}
\author{Daniel Haley\footnote{Corresponding Author: daniel.haley@materials.ox.ac.uk}, Paul A.J. Bagot, Michael P. Moody}
\date{}
\affil{\small{Department of Materials, Oxford University, 16 Parks Road, Oxford, OX1 3PH, UK.}}

\begin{document}
\maketitle

\section{Abstract}
This work describes extensions to existing level-set algorithms developed for application within 
the field of Atom Probe Tomography (APT). We present a new simulation
tool for the  simulation of 3D tomographic volumes, using advanced
level set methods. By combining narrow-band, B-Tree and
particle-tracing approaches from level-set methods, we demonstrate
a practical tool for simulating shape changes to APT samples under applied electrostatic
fields, in three dimensions. This work builds upon our previous studies
by allowing for non-axially symmetric solutions, with minimal loss in
computational speed, whilst retaining numerical accuracy. 
\\
{\bf Keywords:} Atom probe tomography, level set methods, narrow-band, simulation

\section{Introduction}

Atom probe tomography is a 3D characterisation technique for spatially resolving the chemistry of a material at the near atomic-scale. The method works by ionising individual atoms from the apex of a needle-shaped specimen, due to the application of an intense electric field. Ions are subsequently accelerated towards a position sensitive detector placed in the far field. By using time-of-flight mass spectrometry it is possible to determine not only the position, but also the chemistry of these atoms. To generate a 3D tomographic image, it is necessary to use the spatial position on the detector, and the order in which the atoms evaporated to build the 3D location of each atom. The methods that are used to do this are known as reconstruction algorithms, and historically have been geometric in nature~\cite{Geiser2009}. 

It is well known that the geometrical assumption underpinning these reconstruction algorithms is a crude approximation to the physics of the problem~\cite{Oberdorfer2013}, and complex simulations have been undertaken by other authors to simulate a more complete physical model~\cite{Parviainen2015}. The model in wide use today is that of a hemispherical cap, where the mapping between evaporation points on the surface, and the location on the detector is that of a modified stereographic projection.

The simple geometric assumptions in existing reconstruction algorithms are invalid as the materials that evaporate are almost always either inhomogenous or anisotropic. When the tip surface is evaporated and encounters a feature in the tip that evaporates at either a different rate (high or low evaporative field), a protrusion or pit can be created, which causes a lensing effect. Similarly crystalline samples are anisotropic, and evaporate differently along different directions, leading to a non-spherical surface, in contradiction to the projection model's assumption. It is in these nano-scaled inhomogenous and anisotropic samples where atom probe is most useful, and thus it is imperative that the reconstruction be spatially accurate.  This has been identified as a key problem in a range of materials, including semiconductor specimens~\cite{Grenier2014}, oxides~\cite{Mazumder2015} and nanoparticles~\cite{Devaraj2014}. There have recently been new efforts to push the limits of traditional algorithms, using quantitative methods to allow for matching of external data sources (such as Transmission Electron Microscopy) to atom probe datasets~\cite{Mouton2017}. However, even such approaches may not be fully functional in those cases where the projection is subject to strong spatial distortions.

In a previous work, we described the use of so-called ``level set'' algorithms, and the relationship between the problem of field evaporation and the mathematics of level set methods~\cite{Haley2013}. In this prior work, a simple proof-of-concept method was demonstrated for rapid simulation of the field evaporation of atom probe needles incorporating different material phases. Level sets were selected as a mathematical method of encoding geometric information in such a manner that it can be deformed or altered without incurring problems near discontinuities.  In a level set method, the zero-level of the simulation is specified by embedding an $n$-D surface (e.g. a 2D surface) into a $n+1$-D  function. The zero-level here indicates the initial surface of the simulation, and how this is altered through time. By careful construction and manipulation of a 3D function, the 2D surface is moved implicitly. This is highly beneficial over an explicit method, where one needs to resort to ad-hoc methods to remove artifacts caused by motion near such discontinuities.

In atom probe, this allows for the simulation of a tip, evolving subject to field evaporation, given an initial 3D image of the sample. By simulating the velocity of the surface (or equally any point placed at an arbitrary location on said surface), the motion of the tip as a function of time can be computed using the level-set method. It is further possible to modulate the response of the surface to account for any inhomogenous response within the sample.

The field of level-set mathematics has expanded considerably in the last decade, and now has applications from typical multiphase fluid-flow modelling~\cite{Ferrari2017}, to more applied microscopy methods such as the simulation of Focussed-Ion Beam milling~\cite{Bahm2016}. Subsequent to our initial description of the algorithm in the context of atom probe tomography, this has been shown to be of interest to other authors~\cite{Xu2015}~\cite{Bao2015}. Specific case analytical models have also been developed~\cite{Rolland2015}.

The initially proposed model used a low-order approximation to the electrostatic equation, to show that mean-curvature flow (constrained to convex flow) can be used to approximate the shape change enforced on an atom probe needle. This allowed for the use of simple level set models to provide a method of simulating the shape change during an atom probe experiment. However, these models have so far remained as 2D simulations, which simulate either a 2D analogue of the actual 3D system, or otherwise rely on symmetry methods to extend the 2D domain into three dimensions. This however seriously degrades the range of scenarios that can be modelled. For example, a spherical particle that is off the symmetry axis cannot be simulated, as a circle off-axis is a torus under a rotationally symmetric model. As a further example, anisotropies that are not conical in nature cannot be modelled under a cylindrical coordinate system. Thus it is clear that extending the simulation from 2D to 3D is a desirable goal, to allow the simulation model to encompass new behaviour.  3D models that utilise single-atom computations are limited in speed, and can take hours to days to generate a result from a realistically sized needle.

Level set methods allow for the arbitrary spatial scaling, independent of the geometrical precision -- that is, the solution time is not dependent on the absolute scale of the simulation, but rather the resolution of the grid used. Objects at the macroscopic scale, and microscopic scale can be dimensionally rescaled, and provided the physics is still applicable can be modelled with the same computational cost.By utilising voxels of smaller dimension, a more precise refinement of the simulation can be obtained, at a computational cost. Careful selection of computational data structures can alleviate length-scale disparities where small and large objects need to be simulated with differing levels of absolute precision at the same time, thus highly mitigating the computational costs of small inhomogeneities in a larger domain.

High run-times (Circa 12 atoms/min, for a 28~M atom tip~\cite{Oberdorfer2014}, ~1 month for complete evaporation) for these applications mean that performing numerous trial-and-error calculations can be daunting for day-to-day use - however these provide valuable insight into specific effects that can occur within atom probe specimens, such as upon encountering grain boundaries and voids. Earlier attempts to implement 3D simulations were incomplete~\cite{Haley2013}, and had a high computational cost, which scaled poorly with respect to real specimen volumes, rendering the approach of limited utility. The insight gained from rapid feedback in simulations can help atom probe users to better understand and avoid misinterpretation of artefacts within their datasets. We believe such insight can be promoted by the availability of suitable simulation tools, such that developed here.

In this work, we focus on improvements to the representation of level set methods in atom probe, and extend the model from simple 2D axisymmetric solutions to full 3D simulations. This is achieved through the use of narrow-band sparse volume methods, in conjunction with level-set particle tracking for narrow band tracking and reinitialisation for efficient solution of the flow problem. We present a fully functional 3D simulator that provides usable solution times on the order of minutes. We show that the method is sufficiently numerically accurate with respect to theoretical examples of curvature flow for use, and provide sample simulations of 3D tip shape throughout the course of the experiment.

\section{Calculation}

Mean curvature is defined as the average curvatures on a surface in orthogonal directions, i.e. in 3D this is $\kappa = \frac{1}{2} (\kappa_x + \kappa_y)$, and is distinct from Gaussian curvature (product). Hereafter where curvature is discussed, we are referring to mean curvature. As discussed in our previous work, it is possible to show that mean-curvature can be used as a coarse approximation to the action of an electric field on field evaporation~\cite{Haley2013}. Thus mean-curvature flow, in conjunction with level sets can be used to model the tip shape change. Curvature can be derived from a signed distance field, as discussed in detail within the level-set literature~\cite{Sethian2003}, and used to embed an $n$ dimensional surface in $n+1$ dimensional space. In this previous work, 2D simulations (lines embedded in a 2D plane) were shown, as well as axisymmetric 3D simulations (equivalent to surfaces in 3D). A tentative, but computationally inefficient example was shown for 3D.

Here we leverage the OpenVDB framework to obtain higher computational efficiencies and numerical accuracy, in order to allow for the implementation of numerically efficient, scalable simulations of the trajectory of atom probe specimens under an applied electric field. OpenVDB internally utilises a B-Tree type structure~\cite{Cormen04} to group together similar volume data, and to minimise the memory footprint and thus algorithmic operation time for similarly valued voxels, which we leverage for computational efficiency here. 

The simulation consists of several main components, a sparse field system for holding intensity data as a function of position (i.e. the level set values), a spatial discretisation method to obtain curvature from said field, and a temporal integrator to solve the differential equation governing surface flow. Each component is required for the level set framework~\cite{Sethian2003}. In our simulation, the temporal integrator is a simple explicit Runge-Kutta 2nd order solver~\cite{Kreyszig1993}. Timesteps are dynamically computed to ensure that the Courant–Friedrichs–Lewy (CFL) criterion is met, up to constant factor $C$. We use the OpenVDB~\cite{Museth2013} framework to provide a sparse simulation framework, and to provide level set initialisation. This allows us to use so-called ``narrow-band'' level sets, which provide significant speed increases in solving the differential equation, by limiting the computational domain. Temporal integration of the grid, and curvature computation is performed using central-difference gradient operations to compute the mean curvature field, using second-order gradient operators. 

Particles are injected into the simulation to implement a so-called ``particle-level set'' method~\cite{Losasso2006}, whereby particles are initially seeded on an isosurface of the initial zero level. This is done to enable quantitative tracking of the motion of an unconstrained tracer within the field, as compared to the underlying discretised grid. In a non-discretised case, the particle and zero-level should perfectly track one another, however in the case of a discretised volume errors occur as deviations in the zero level and the relative position of these particles as time progresses~\cite{Enright2005}.  Thus the use of particles allows reinitialisation as often or little as needed, and thus improves accuracy by minimising numerical dissipation.

These particles are then moved by the curvature of the field in the same manner as the signed distance is modified, but with a vector of curvature, rather than its norm. The particle trajectories are then temporally integrated using a first-order Euler integrator to move them, and provide an estimate of the new zero level's position. Here, if any of this set of particles approach the full width of the narrow band, to within a tolerance factor, $0 < \beta < 1$, then the signed distance field is reinitialised from the zero level isosurface, and the particles reset to the new surface. Without this reinitialisation, the surface is unable to escape the initial narrow band, and thus cannot follow its correct trajectory.

\section{Results}
\subsection{Numerical testing}
A simple, and common test for numerical accuracy in level set methods is to check the rate at which a sphere collapses under its own mean curvature. A sphere has constant mean curvature, inversely proportional to the radius, $r$:

\begin{equation}
\kappa =  1/r
\end{equation}

When solving the mean curvature flow equation, we solve the change in the surface over time, related to curvature by $\frac{dr}{dt} = -k$, enforcing that curvature should be positive to prevent growth of concavities, which is nonphysical in an atom probe context. As per the level set method, we construct an initial signed distance field, over the solution domain $\phi$, such that the value of $\phi$ is defined by an initial level set. Note that at $t=0$, the solution appears to have ``terraces'' - these are simply as a result of the binarisation of an image to create the initial sphere, and are almost immediately smoothed out by the curvature operator - these do not represent a specific material or crystal structure.

Solving this for a sphere of initial radius $r_0$, we obtain that the sphere's radius over time, $r(t)$,  will be:
\begin{equation}
r(t) = \sqrt{r_0^2 -t}
\end{equation}

\begin{figure}[hp]
 \centering
 \includegraphics[width=0.95\textwidth]{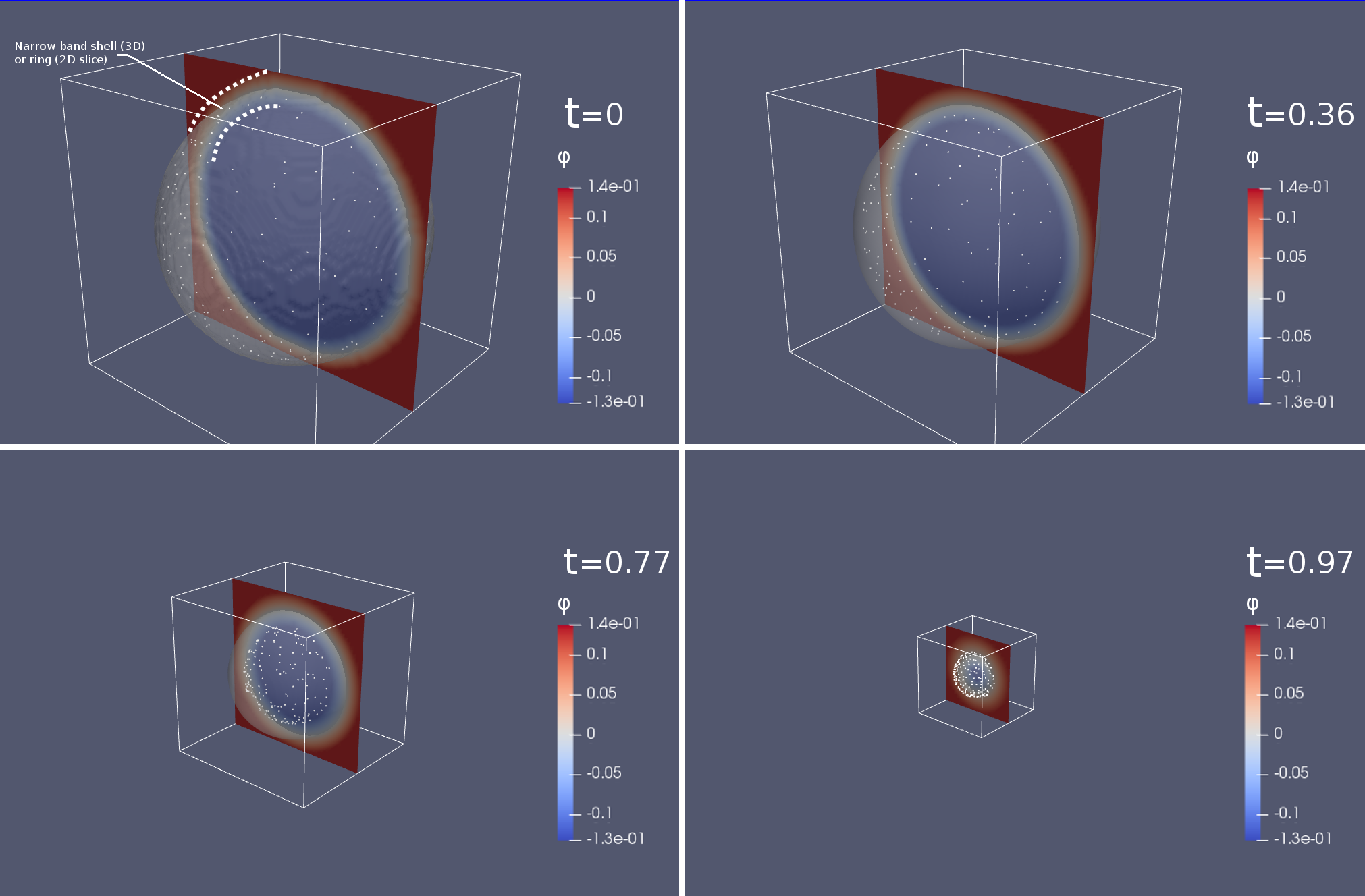}
 \caption{Sphere-collapse test: A unit sphere should collapse under its own curvature, and arrive at zero radius at $t=1$. The isosurface shows the zero-level surface (i.e.\ the interface we wish to track). Particles show the tracers used to maintain the sparse grid. The spatial scale is constant in all frames - the computational domain is seen to dynamically shrink. Total simulation time is 10.9~s. Roughness in initial frame is due to binarisation of the initial data used to generate the initial $\phi$, and should be ignored.}
 \label{fig:sphereCollapse}
\end{figure}

\begin{figure}[htp]
 \centering
 \includegraphics[width=0.95\textwidth]{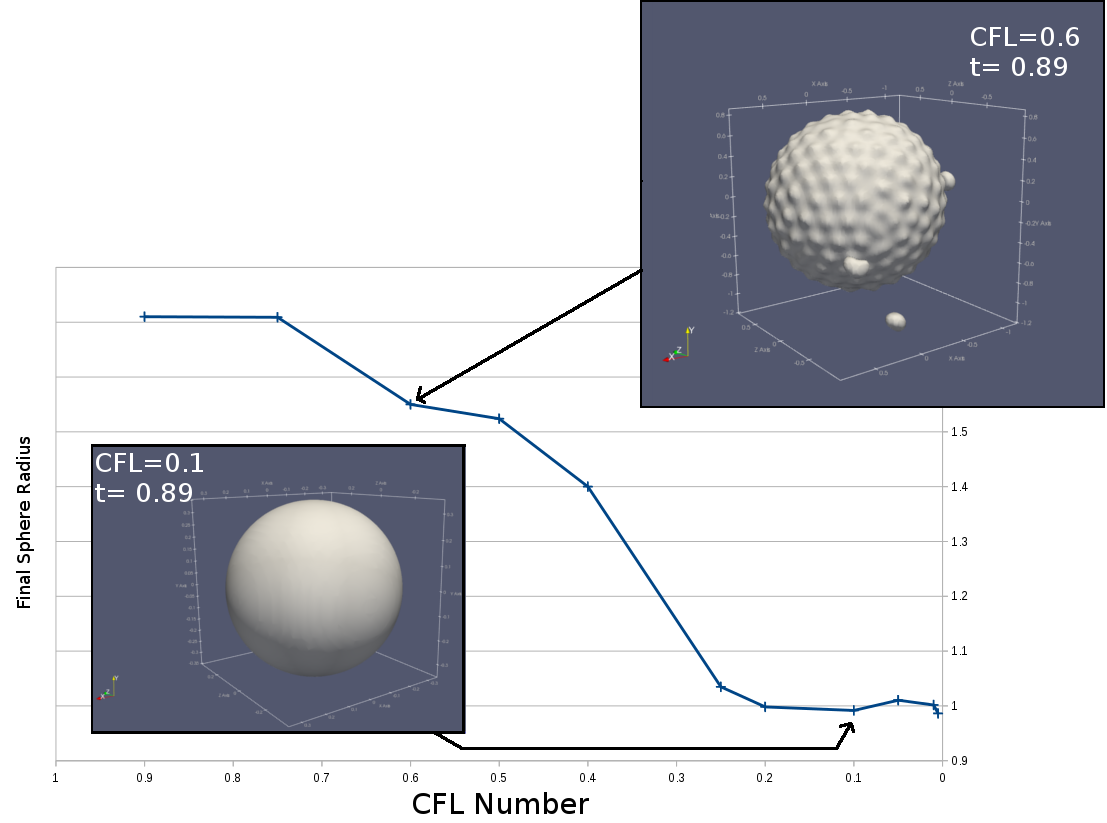}
 \caption{Demonstration of convergence as a function of CFL number.  Simulation is unstable for $\mathrm{CFL}  > 0.25$, resulting in visible instabilities in the surface of the simulation. }
 \label{fig:CFLConvergence}
\end{figure}

Where $r_0$ is the initial sphere radius. Figure~\ref{fig:sphereCollapse} shows the numerical solution of this using our simulation, solving for a sphere ($r_0=1$), where the sphere self-collapses at $t=1$, demonstrating that the simulation satisfies this test to an acceptable level, with the numerical estimate of $t=1.00168$, rather than 1. The real time required for the simulation on an i5-4670 CPU (3.40GHz) is 10.9~s (narrow band width=6, max time step=0.1, $C=0.01$, initial voxels/side = 80), where $\mathrm{CFL}$ is a tolerance coefficient (the CFL number) that reduces the timestep dynamically, based upon the maximum curvature value, such that the timestep $dt$ is governed by the curvature and the grid size $dx$~\cite{Wang2004}. Figure~\ref{fig:CFLConvergence} shows the change in the final radius and time for the sphere collapse test as a function of $\mathrm{CFL}$.

\begin{equation}
 dt = \min \left(\frac{C}{3} \frac{dx}{\max(|\kappa|)},t_{\mathrm{max}}\right)
\end{equation}

In Figure~\ref{fig:sphereCollapse}, the narrow ring shown in the plane is the solution domain, and no computations are shown outside that domain - including in the interior of the sphere. Indeed, only a thin shell is simulated, which provides significant computational efficiency. Periodic reintialisation of the level set is required, as the surface moves towards the edge of the initial sparse domain. As can be seen, the simulation maintains close to the correct trajectory for the majority of the simulation time, deviating near the end of the simulation, due to loss of grid resolution. As this simulation is a numerical approximation to the solution of the differential equation, if the convergence parameters are not correctly chosen, then effects such as ``drift'' from the solution may be apparent. 

To obtain a solution for the radius, the zero-level contour is extracted using an isosurface algorithm provided by OpenVDB. The mesh nodes $p_i$ from this isosurface are used to estimate the the radius for the sphere.

\begin{equation}
 R' = \displaystyle \sum_{i=1}^n ( || p_i - u ||)
\end{equation}

Where $u$, the centre of the sphere, is given by the particles ($p_i$) in the cloud's centroid:

\begin{equation}
 u = \frac{1}{n} \displaystyle \sum_{i=1}^n p_i
\end{equation}

\begin{figure}[htp]
 \centering
 \includegraphics[width=0.95\textwidth]{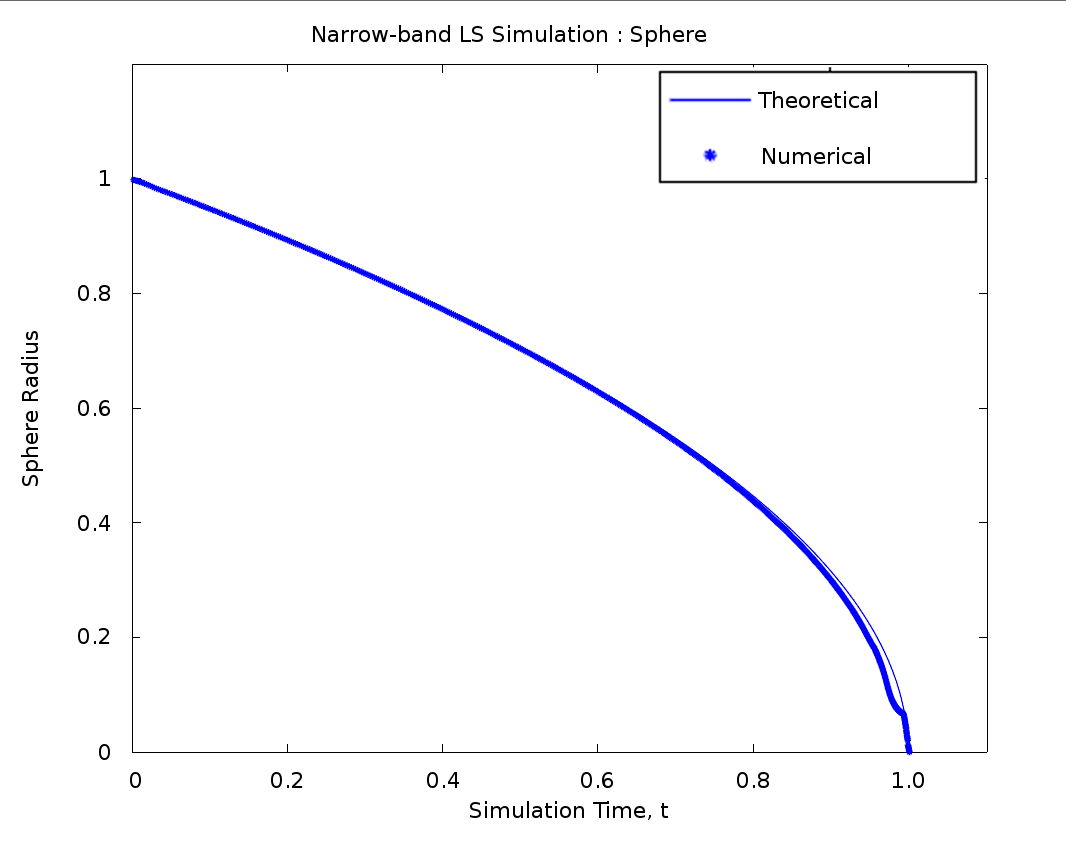}
 \caption{Radius as a function of time, showing that the theoretical collapse time (solid line) is reasonably well matched to the numerical solution (markers), up until the sphere is small as compared to the grid size.}
 \label{fig:sphereCollapseTime}
\end{figure}

The trajectory is shown in Figure~\ref{fig:sphereCollapseTime}. Here it is clear the trajectory is reasonably well approximated to the theoretical curve, indicating that the program is solving the problem without considerable numerical issues. 

There are some deviations noticeable however, where there exists a small deflection from the theoretical curve at $t \approx 0.85$, possibly due to numerical diffusion during the Gaussian smoothing of the mean curvature, ``drift'' of the level set (where $||\nabla \phi|| \neq 1$), or that the initial approximation was based upon a discrete approximation to the sphere. A further unphysical deflection occurs at $t \approx 0.97, r \approx 0.1$, where the sphere radius is now comparable to the grid size -- $\frac{0.1}{0.0125}=8~\mathrm{voxels/side}$. The sphere curvature becomes undefined as $r \rightarrow 0$, and the simulation starts to become unstable as the sphere can no longer be well approximated by the grid.

Lastly, the spherical solution shows an important, and well known, feature of curvature flow -- self similarity. This is where the temporal evolution of a sphere under curvature flow is itself a sphere, which implies that any shape that becomes a sphere will stay in that form until it collapses. This is an important result for the equilibrium shape of atom probe tips, and indeed extends from sphere to ellipsoidal objects, and other initially convex objects~\cite{Olver1997}, which can flow to a self similar end-result (not necessarily the initial shape). There is a strong relationship between classical geometric atom probe reconstruction, and self-similarity - the equilibrium form may not affected by tip size, and there is an implication that there may be a continuous, smooth, and possibly self-similar transformation from the initial to the final shape. A detailed analysis of the nature of this relationship is outside the scope of this work.

\subsection{Tip simulation}
\subsubsection{Homogenous simulation}
\begin{figure}[htp]
 \centering
 \includegraphics[width=0.95\textwidth]{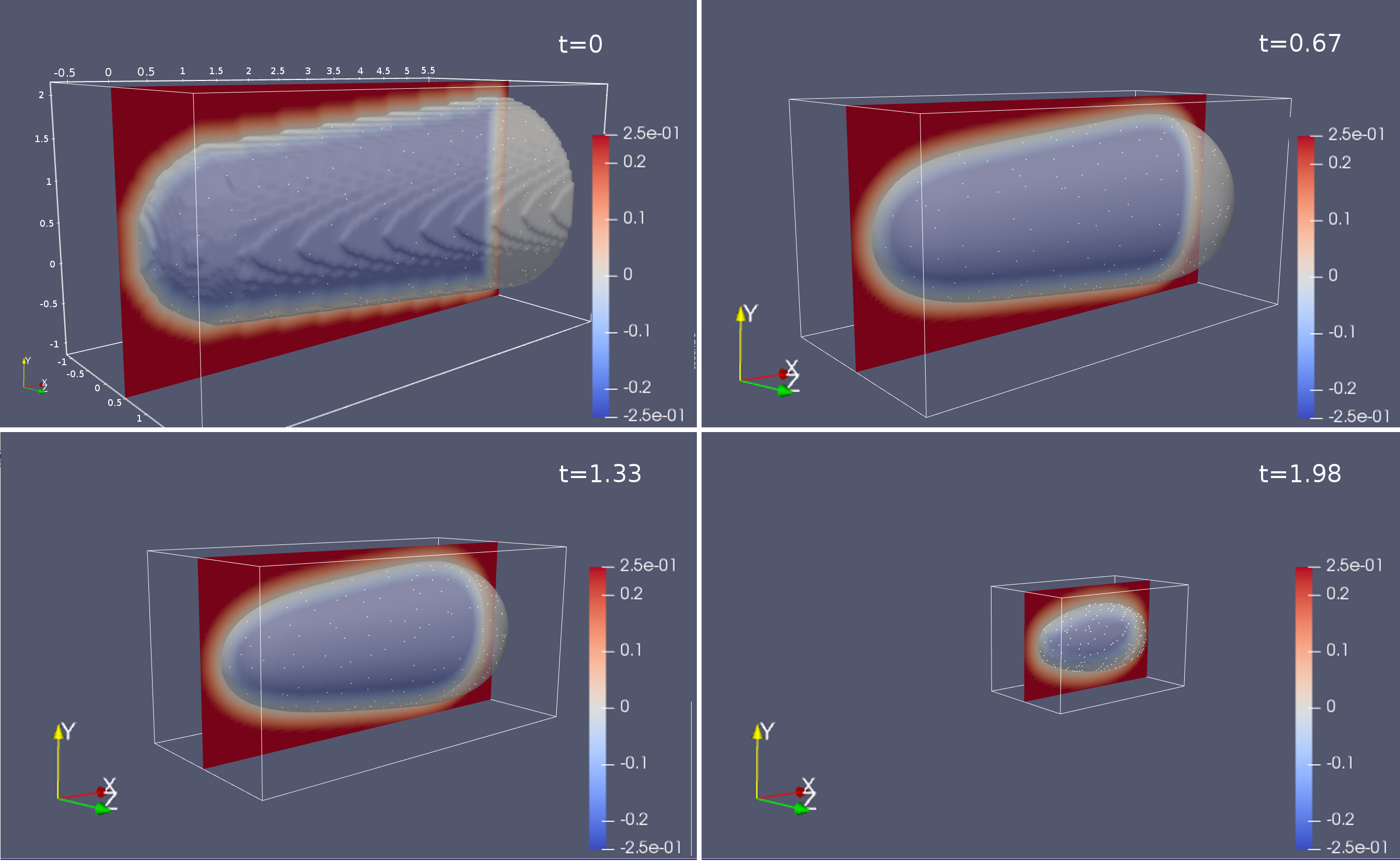}
 \caption{Simulation of truncated tip without imposed boundary conditions, undergoing curvature flow. Scale is identical in all frames, settings are as per the sphere simulation.  Note how the truncated end of the tip is shrinking, which is a nonphysical result observable in our previous work~\cite{Haley2013}.}
 \label{fig:tipfreesimulation}
\end{figure}

As per our previous work~\cite{Haley2013}, we demonstrate a hypothetical simulated tip, and show the evolution of the tip through time. In this simulation (Figure~\ref{fig:tipfreesimulation}), no boundary conditions are imposed, and as such the hypothetical tip is evaporated, but converges to an ``egg-like'' shape, with smoothly differing curvatures across the surface. The settings used in the temporal evolution, and spatial discretisation are the same as the sphere algorithm (excepting the presence of the shank), however the spatial scale is altered (which does not alter the shape trajectory). The total time required for the simulation is 121~s.

\begin{figure}
 \centering
 \includegraphics[width=0.95\textwidth]{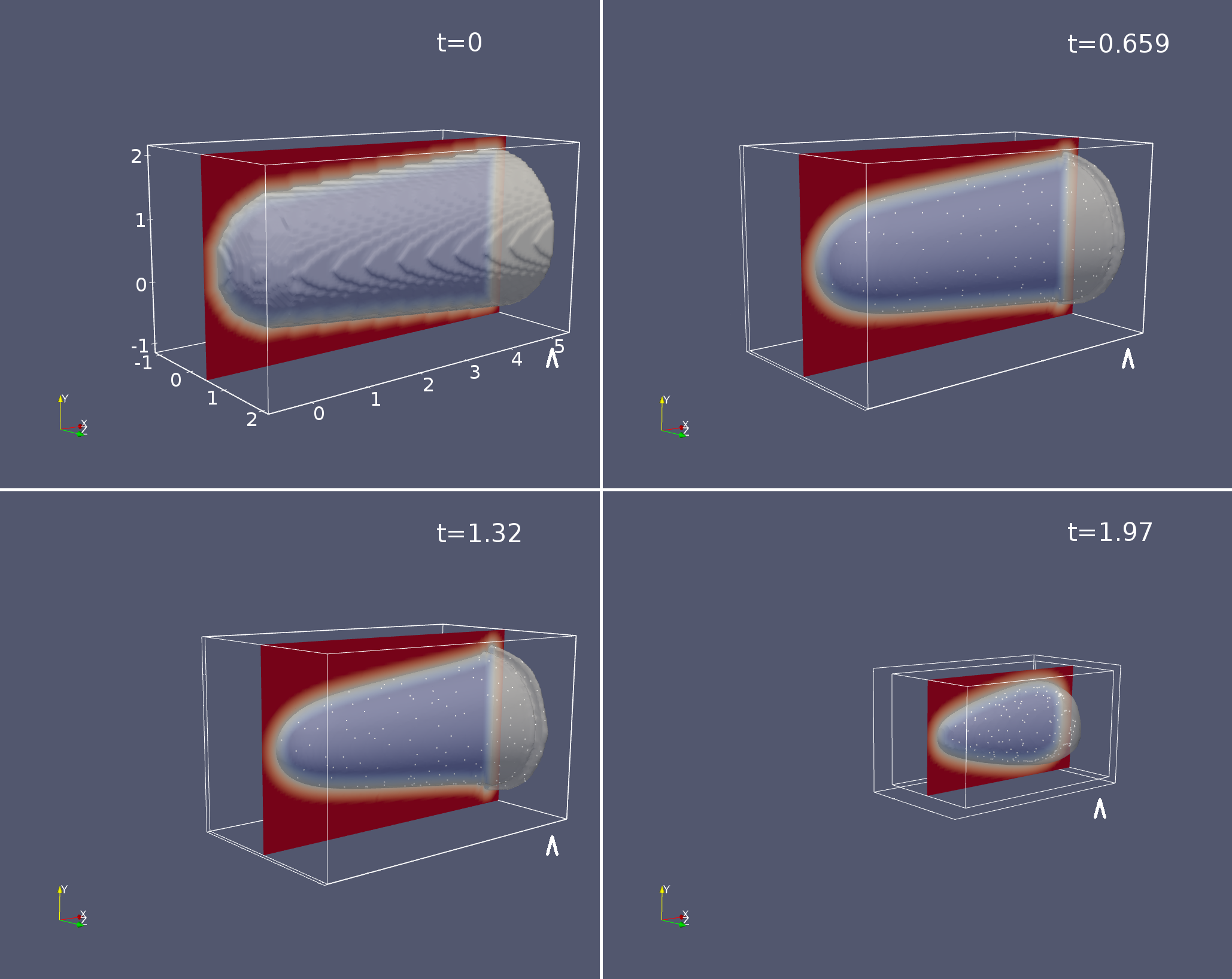}
 \caption{Simulation of truncated tip fixed boundary condition at the truncated end, undergoing curvature flow, total time to complete simulation is 213~s. In this case, the tip correctly shrinks at only one end, however numerical instabilities are visible near the truncated end, and the sides of the tip are removed too quickly as compared to an atom probe experiment, due to the curvature approximation to the Poisson equation.}
 \label{fig:tipfixed}
\end{figure}

To overcome this nonphysical evaporation, we utilise ``fixed'' nodes at the truncated point on the specimen. These are assigned a zero curvature during evolution regardless of the level-set field. This has the effect of preventing evaporation at the position where the nodes are specified. In the actual implementation this is performed by creating a second grid of differing dimension, where the boundary nodes are assigned a zero value, and then multiplying the computed curvature from the original grid with this. The results of this simulation are given in Figure~\ref{fig:tipfixed}.

\subsubsection{Inhomogenous simulation}
The two previous simulations show how a tip composed of a homogenous, isotropic material will dynamically change shape over time. However, real material problems are almost invariably for inhomogenous systems. As per the previous work, the effect of differing phases with an inhomogenous evaporative field is modelled by using a second image, $I$, to multiply the curvature value, to change the effective evaporation rate. Reducing the curvature ($I(\overline v)<1$) simulates materials that evaporate at a higher field than the surrounding material, whereas increasing the curvature ($I(\overline v)>1$) causes the material to have an effectively lower evaporation rate.

The simulation here shows the effects of this inhomogenous evaporation, specifically one that does not have axial symmetry. This simulation cannot be reproduced using an axially symmetric model. In this simulation, the lower half of the tip has the effective curvature modified by a 3D image, $I'$, where the curvature is reduced by a factor of 0.2. $I'$ is generated from the fuction $I'(x,y<=0.5,z) = 0.2, I'(x,y>0.5,z) = 1$.

\begin{figure}[ht]
 \centering
 \includegraphics[width=0.95\textwidth]{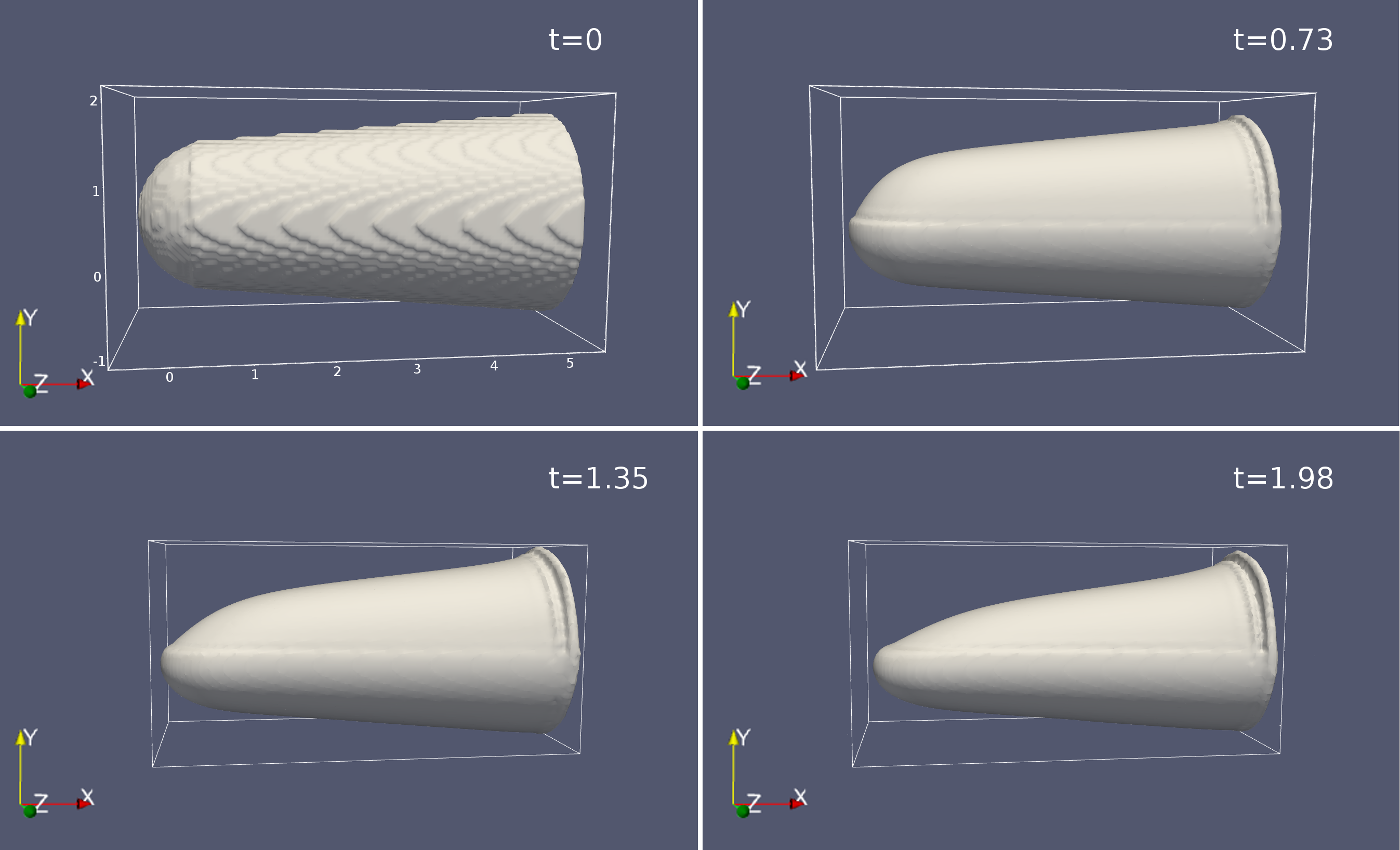}
 \caption{Simulation of truncated tip, with fixed boundary, undergoing curvature flow, where the lower half is modified by reducing the velocity of the surface by multiplying the geometric curvature by a factor of 0.2. This scenario cannot be simulated in an axisymmetric simulation. Total execution time is 42~s.  Visible artefacts occur at truncated zone, however the leading section of the tip correctly reduces in the expected manner for curvature flow.} 
 \label{fig:tiphalf}

\end{figure}
\section{Discussion}

\subsection{Prediction errors}
Whilst the simulation shown is accurate in a mean-curvature sense, the simulation does not provide a fully coupled numerical solution to field-evaporation, as it only provides a first-order approximation to the electric field. 

There are several major effects where this is clear - firstly, it can be seen during the tip-test, that the sides of the specimen evaporate more than is physically realistic. The source of this is two fold - firstly the curvature at these regions is nonzero, but the flat direction of the tip does not provide any decrease in the electric potential gradient, as would be expected. More accurate solvers, which allow for a computational speed-accuracy tradeoff by iteratively solving the Poisson equation will resolve this concern. Secondly, a smaller effect where the simulation is stabilised by smoothing the curvature before temporal integration. We believe this step should not be required, however attempts to solve the simulation without this resulted in an unstable simulation. Switching to electrostatic solutions should again eliminate the need to artificially smooth the temporal update value, as smoothing will be inherent in the summation that provides the update value.

Indeed, we here state that we have developed 2D simulations which do not rely on the curvature assumption, but rather perform a coupled electrostatic algorithm, which will be the subject of a future publication. The use of an electric field implies that we can only modify the level-set field below the zero-level (within the vacuum), as within a conductor the field is zero. We can however, maintain accuracy using a coarse approximation to the future electric field by extending the electric field values slightly into the internal grid, and using curvature elsewhere.

Such an approach is crude, but will yield a more correct evaporation form for the tip, which is more physically realisable. The alternative is to perform reinitialisation more regularly, to minimise errors - however this has a moderately increased computational cost. We are currently exploring the use of velocity extensions~\cite{Chopp2009}, which we believe will eliminate these concerns and provide very acccurate simualations, without any major alternations to the surrounding level-set method as presented here.

\subsection{Framework concerns}

Boundary effects are not sufficiently handled here, leading to a slight numerical diffusion of intensity (material) within the simulation volume, or artefacts around fixed boundary conditions. Improving boundary handling will reduce the inaccuracy, ensuring that material volume is removed in an accurate manner. Higher order ``Total-Variation-Diminishing'' integrators may also aid numerical accuracy whilst allowing for larger time steps, or an increase in $C$, which will significantly enhance the time/accuracy of the simulation, enabling solutions to be accurately obtained more rapidly.  

There exist numerous avenues to further improve the performance of the calculation at a programmatic level. At the simplest level, there are several data copy operations during temporal integration that can be avoided. Whilst the implementation is currently parallelised for a single host, there is considerable scope for improvement in parallelism. Parallel-in-time integrators do exist~\cite{Emmet2012}, which allow for continuously improving the time resolution of a simulation, and as such can readily be distributed across multiple systems. This would have the advantage of further accelerating the simulation and test phase for a full reconstruction process. Automatic determination of shape convergence would also be highly valuable, to allow the user to specify a maximum discrepancy in surface shape. This could be coupled with post-APT electron microscopy examination, to provide boundary variables for ensuring accurate shape estimation.

\subsection{Simulation Code}
The simuation code is available at \url{http://apttools.sourceforge.net} under the GNU General Public Licence (v3+). 

\section{Conclusions}

We have developed the first fully 3D level-set based sparse method for atom probe tomography. The combination of level sets and spatially sparse solvers allows for numerically accurate calculations to be performed in 3D quickly. The 3D nature of these simulations allows for non-axisymmetric problems to be tackled in a numerically robust, and rapid fashion. Such problems are the mainstay of real atom probe datasets, where the geometry of the tip is governed by the physical material of interest. We have shown that there is no computational impediment to the implementation of the simulation of a smoothly transitioning surface by curvature flow, as simulations can be executed in minute timescales on standard desktop computers.

More advanced simulations implementations, extended from the framework shown here,  may form the cornerstone of future APT simulation tools. We believe the integration of such solutions with a reconstruction process is now within reach, and this may well provide a practical route to a fully physically motivated reconstruction within usable timescales to make it of benefit to real atom probe experiments. 

These simulations will have the potential to enable physically accurate reconstruction, where geometric distortions introduced by anisotropies or inhomogeneities in the sample can be fully accounted for, and provide the theoretical framework necessary for eliminating spatial distortions within atom probe datasets.

\section{Acknowledgements}

We wish to acknowledge the EPSRC Hems project EP/L014742/1.

\section {Publication}
This has been published in Materials Characterization, 10.1016/j.matchar.2018.02.032.

\bibliographystyle{unsrt}
\bibliography{levelset_3d.bib}
\end{document}